\documentclass[twocolumn,showpacs,preprintnumbers,amsmath,amssymb]{revtex4}

\usepackage{graphicx}
\usepackage{dcolumn}
\usepackage{bm}
\begin{document}

\title{Umklapp scattering and electron pairing cutoff in BCS superconductors}
\date{\today}
\author{X.\ H.\ Zheng and D.\ G.\ Walmsley}
\affiliation{Department of Pure and Applied Physics,
The Queen's University of Belfast,
Belfast BT7 1NN, Northern Ireland}
\email{xhz@qub.ac.uk}

\begin{abstract}
In a superconductor electrons form pairs for which the end states
of normal and umklapp scattering may overlap.  This cuts electron
pairing off at a phonon frequency, $\omega_c$, low compared with
the Debye frequency, $\omega_D$.  Using this insight, correct
values of $2\Delta/k_BT_c$ (average error 8.9\%) for 12
superconductive metals, including Hg and Pb, are achieved from
simple BCS formalism with an average $\omega_c/\omega_D$ of 0.148:
Landau's idea of a Fermi liquid may cover strong-coupling
superconductors.  The cancellation between normal and umklapp
scattering may be more important than a strong electron-phonon
interaction in reaching a high critical temperature $T_c$.
\end{abstract}

 \maketitle

\begin{figure}
\resizebox{8cm}{!}{\includegraphics{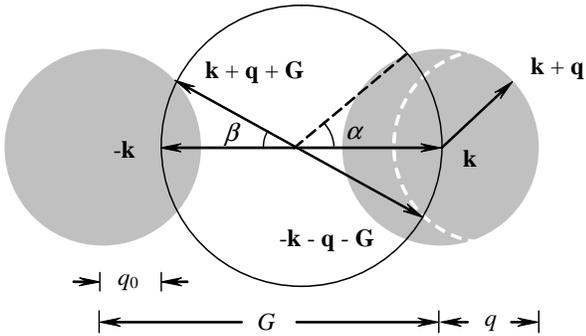}}
\caption{\label{fig:fig1} A spherical Fermi surface (open circle)
and a pair of electrons in states ${\bf k}$ and ${\bf -k}$. This
pair is scattered into states ${\bf k + q + G}$ and ${\bf -k - q -
G}$ in the umklapp process, ${\bf q}$ being the phonon wavevector,
${\bf G}$ the reciprocal lattice vector. The shaded circles
represent two identical spherical phonon zones, angles $\alpha$
and $\beta$ measure the size of the sections of the Fermi surface
being intersected by these phonon zones.  We have $\alpha
> \beta$ because a larger section of the Fermi surface is
intersected by the phonon sphere centered at this surface (the
other sphere is centered above the Fermi surface at height $q_0$).
As a result, the electron state ${\bf -k - q - G}$ must lie inside
the phonon sphere centered at ${\bf k}$, where all the states are
involved in normal scattering. Furthermore, when ${\bf k + q + G}$
runs over the shaded phonon zone on the left, ${\bf -k - q - G}$
also runs over a sphere indicated in part by the broken white
circle (centered above the Fermi surface at height $q_0$ to mirror
the phonon sphere on the left).}
\end{figure}

Despite the enormous effort in nearly 50 years since
BCS~\cite{BCS}, a fundamental issue at the very heart of the study
of superconductivity appears to have escaped attention: in a
superconductor the end states of normal and umklapp scattering may
overlap (FIG.~\ref{fig:fig1}).
The resulting competition for end states frustrates the electron
pairing process and leads to a phonon cutoff frequency,
$\omega_c$, which was postulated by Bardeen, Cooper and Schrieffer
(BCS) but not specified~\cite{BCS}, and is often taken to be the
Debye frequency, $\omega_D$, in the current
literature~\cite{Ketterson}.  However $\omega_c$ must be low to
reflect the low onset frequency of umklapp scattering.  From
numerical solutions of the BCS self-consistent equation, an
average $\omega_c/\omega_D$ of 0.148 is required to reproduce
experimental values of the energy gap, $\Delta$, for 12
superconductive metals, Hg and Pb included.  Without considering
${\bf k}$-dependence, the low $\omega_c$ leads to substantially
correct values of $2\Delta/k_BT_c$ (average error 8.9\%,
Table~\ref{tab:table1}, FIGs.~\ref{fig:fig2} and~\ref{fig:fig3}).
It therefore appears that simple BCS theory, based on Landau's
idea of a Fermi liquid, is applicable to at least some
strong-coupling superconductors.  Hence pairing frustration
arising from the competition between normal and umklapp scattering
may be more important than a strong electron-phonon interaction in
reaching a high $T_c$.

First we see how in a metal the onset frequency of umklapp
scattering is usually much lower than $\omega_D$.  Consider the
spherical Fermi surface in Fig.~\ref{fig:fig1} that does not
intersect the boundary of the Brillouin zone: there is a minimum
distance $q_0$ between adjacent Fermi surfaces.  We can prove by
geometry that $q_0/q_D = 0.02$ for the sc lattice with one
electron orbital per atom, or $q_0/q_D = 0.28$ for the bcc lattice
with two electron orbitals per atom, $q_D$ being the Debye
momentum. If the phonon momentum $q$ exceeds $q_0$ then an
electron in state ${\bf k}$ may be scattered into ${\bf k + q}$ to
lie on the adjacent Fermi surface. It makes no physical difference
if the reciprocal lattice vector ${\bf G}$ (parallel to ${\bf k}$
for simplicity) is added to an electron state: we can replace
${\bf k + q}$ with ${\bf k + q + G}$, which lies on the original
Fermi surface, and is known as the end state of umklapp
scattering; it is usually far away from the initial state of the
electron.

Umklapp scattering often plays a significant role in solids at low
temperatures~\cite{Ziman}, and superconductors may be no
exception. Let ${\bf k}$ and ${\bf -k}$ be the states of a Cooper
pair, which by nature are on opposite sides of the Fermi sea, like
a pair of rotating doors. Correspondingly ${\bf k + q + G}$ must
be paired with ${\bf -k - q - G}$: scattering of the electron pair
is synchronized, reminiscent of the synchronized motion of
rotating doors.  It is evident from Fig.~\ref{fig:fig1} that ${\bf
-k - q - G}$ lies inside the phonon sphere centred at state ${\bf
k}$, where all the electron states are involved in normal
scattering: thus the end state of umklapp scattering may not be
empty, a situation that may have serious consequences.

In order to incorporate umklapp scattering into the BCS formalism,
we must make sure that in this formalism normal and umklapp
scattering do not drive electrons into the same end state at the
same time.  We let
\begin{eqnarray}
&& |\Psi\rangle = \prod _{{\bf k}(\mathcal N)}
[(1-h_{\bf k})^{1/2}\nonumber\\
&& \;\;\;\;+ h^{1/2}_{\bf k}b^+_{\bf k}]\prod _{{\bf k}(\mathcal
U)} [(1-h_{\bf k})^{1/2}b^+_{\bf k} - h^{1/2}_{\bf k}]|0\rangle
\label{eq:ground}
\end{eqnarray}
be our ground state wave function, where $b^+_{\bf k}$ is the pair
generation operator, $h_{\bf k}$  occupation probability,
$|0\rangle$ the vacuum, $\mathcal N$ and $\mathcal U$ specify
paired end states of normal and umklapp scattering (or normal and
umklapp pairs for short) respectively.  It is easy to check that
wave functions of the normal and umklapp pairs with the same ${\bf
k}$ are always orthogonal to each other: the probability is zero
for normal and umklapp scattering to share the same end state at
the same time. The ground state energy $W =
\langle\Psi|H_{\hbox{\scriptsize BCS}}|\Psi\rangle$ is minimized
if $\partial W/\partial h_{\bf k} = 0$, $H_{\hbox{\scriptsize
BCS}}$ being the BCS reduced Hamiltonian \cite{BCS}, which leads
through the constraint $h_{\bf k} = (1/2)[1 - \epsilon_{\bf
k}/E({\bf k})]$ to
\begin{equation}
\Delta({\bf k})= \biggl[\;\sum_{{\bf k'}(\mathcal N)} -\sum_{{\bf
k'}(\mathcal U)}\;\biggr]V_{\bf kk'}\frac{\Delta({\bf k'})}
{2E({\bf k'})} \label{eq:self}
\end{equation}
where $E({\bf k}) = [\Delta^2({\bf k}') + \epsilon^2_{\bf k'}]^2$,
$\epsilon$ being the energy relative to the Fermi level, ${\bf k}$
can be specified by either $\mathcal N$ or $\mathcal U$.

According to the above self-consistent equation normal pairs
sustain superconductivity but umklapp pairs frustrate it.  This is
true only when the normal pairs dominate, i.e.\ some terms in
${\bf k'}(\mathcal N)$ survive cancellation by terms in ${\bf
k'}(\mathcal U)$.  Otherwise $\partial W/\partial h_{\bf k} = 0$
will become the condition to maximize $W$, unless we choose
$h_{\bf k} = (1/2)[1 + \epsilon_{\bf k}/E({\bf k})]$ as the
constraint and consequently, we have to exchange $\mathcal N$ and
$\mathcal U$: then umklapp pairs sustain superconductivity but
normal pairs frustrate it.  It appears that either normal or
umklapp pairs alone may lead to an energy gap ($h_{\bf k}$ or $1 -
h_{\bf k}$ interpreted as the occupation probability
respectively), but their coexistence mutually cancels their
effect.  We are reminded that $(1 - h_{\bf k})^{1/2}b^+_{\bf k} -
h^{1/2}_{\bf k}$ was also used by BCS to generate excited pairs,
apparently in order to forbid excited and ground pairs sharing the
same state at the same time, so that the entropy of the particle
ensemble could be counted properly~\cite{BCS}.  No confusion need
arise. If the normal pairs dominate, then all the terms in ${\bf
k'}(\mathcal U)$ will be cancelled: excited pairs will be
introduced in the absence of umklapp pairs.  If the umklapp pairs
dominate, then we use $(1 - h_{\bf k})^{1/2} - h^{1/2}_{\bf
k}b^+_{\bf k}$ to generate excited pairs. To make the discussion
concrete we assume dominance of normal pairs.

A note about the Coulomb repulsion, which was dropped from the
reduced Hamiltonian by BCS~\cite{BCS}.  According to calculations
based on the Bogliubov model potential, the Coulomb repulsion has
negligible effect on the value of the energy gap in the BCS
theory, compared with the effect of the electron-phonon
interaction, if the so-called Coulomb cutoff frequency, introduced
artificially, is large enough.  In fact, an energy gap may arise
even from an entirely repulsive interaction, when this
inter-action is perturbed by an attractive interaction over a
narrow range of phonon frequencies~\cite{Ketterson}.  It appears
that in a metal Coulomb repulsions between electrons are in a
balance which can be toppled even by a weak attractive
interaction.

We may also justify the reduced Hamiltonian from the
field-theoretic point of view.  This Hamiltonian arises from the
canonical transformation $H_{\hbox{\scriptsize BCS}} =
e^{-S}He^S$, where $H = H_{\hbox{\scriptsize e}} +
H_{\hbox{\scriptsize p}} + H_{\hbox{\scriptsize e-p}}$ with
$[H_{\hbox{\scriptsize e}} + H_{\hbox{\scriptsize p}}, S] +
H_{\hbox{\scriptsize e-p}} = 0$, $H_{\hbox{\scriptsize e}}$,
$H_{\hbox{\scriptsize p}}$ and $H_{\hbox{\scriptsize e-p}}$ are
Hamiltonians of electron, phonon and electron-phonon interaction,
respectively, the brackets represent the operation $[A, B] = AB -
BA$.  With this canonical transformation we cancel
$H_{\hbox{\scriptsize e-p}}$ in the first order expansion of
$H_{\hbox{\scriptsize BCS}}$\cite{Kittel}.  Apparently we can do
the same when $H = H_{\hbox{\scriptsize e}} + H_{\hbox{\scriptsize
p}} + H_{\hbox{\scriptsize col}} + H_{\hbox{\scriptsize e-p}}$
(i.e. to use $H_{\hbox{\scriptsize e}} + H_{\hbox{\scriptsize p}}
+ H_{\hbox{\scriptsize col}}$ to replace $H_{\hbox{\scriptsize e}}
+ H_{\hbox{\scriptsize p}}$ in the above canonical transformation,
$H_{\hbox{\scriptsize col}}$ being the Hamiltonian of the Coulomb
repulsion), provided that we use the eigen-functions of
$H_{\hbox{\scriptsize e}} + H_{\hbox{\scriptsize p}} +
H_{\hbox{\scriptsize col}}$ (instead of the eigen-functions of
$H_{\hbox{\scriptsize e}} + H_{\hbox{\scriptsize p}}$) as the base
functions in the second quantization.  Now $H_{\hbox{\scriptsize
BCS}}$ measures the second order perturbation to the energy of the
particle ensemble $with$ the Coulomb repulsion. This brings little
change to the BCS theory, except that in $V_{\bf kk'}$ the
electron energy may be shifted slightly.

Let us proceed.  In principle we can identify the terms in ${\bf
k'}(\mathcal N)$ that have survived cancellation by terms in ${\bf
k'}(\mathcal U)$, or vice versa, and go on to solve the
self-consistent equation, a daunting task due to the incompatible
symmetries of the Fermi sea and the reciprocal lattice.  It is
clear from Fig. 1 that we have to find ${\bf k'}(\mathcal N)$ and
${\bf k'}(\mathcal U)$ all over again whenever the direction of
{\bf k} changes.  In order not to do so we follow BCS and assume
that phonons are cut off abruptly at $\omega_c$. We also assume an
isotropic energy gap function written as $\Delta(\epsilon)$. We
will vary {$\omega_c$ to let the calculated $\Delta$ match its
experimental value. Our theory will be supported if
$\omega_c/\omega_D << 1$ holds to reflect the low onset phonon
frequency for umklapp scattering.

We adopt the free electron model, on the basis that the BCS theory
is based on the principle of variation, which is not sensitive to
errors in the trial function.  We also adopt the Debye phonon
model.  Therefore in
\begin{equation}
V_{\bf kk'}= \frac{2\hbar\omega _{\bf q}|\mathcal{M}_{\bf q}|^2}
{(\hbar\omega_{\bf q})^2 - (\epsilon_{\bf k'} - \epsilon_{\bf
k})^2}
\label{eq:V}
\end{equation}
where $\omega_{\bf q}$ is the phonon frequency, ${\bf q = k -
k'}$~\cite{BCS}, we have $\epsilon_{\bf k'} = \epsilon_{\bf k} +
4\epsilon_F\xi x$ for states close to the Fermi surface,
$\epsilon_F$ being the Fermi energy, $\xi = q/2k_F$, $x =
\cos\theta + \xi$, $k_F$ the Fermi wavenumber, $\theta$ angle
between ${\bf k}$ and ${\bf q}$.  Consequently we have
$(\hbar\omega_{\bf q})^2 - (\epsilon_{\bf k'} - \epsilon_{\bf
k})^2 = (4\epsilon_F\xi)^2(\delta^2 - x^2)$, $\delta =
\hbar\omega_{\bf q}/4\epsilon_F\xi = (1/4\xi)(\omega_{\bf
q}/\omega_D)(\hbar\omega_D/\epsilon_F) = (k_F/2q)(q/q_D)(T_D/T_F)
= (Z/16)^{1/3}(T_D/T_F)$, $T_D$ and $T_F$ are Debye and Fermi
temperatures.  We have $\delta\sim 10^{-3}$ in
Table~\ref{tab:table1}, but in more realistic models $\delta$ may
not be a constant.
\begin{table}
\caption{\label{tab:table1}}
\begin{ruledtabular}
\begin{tabular}{ccccccc}
 & $2\Delta _0$\footnotemark[1]
 & $2\Delta(0)$\footnotemark[1]
 & $\delta$\footnotemark[2]
 & $\omega_c/\omega _D$
 & $2\Delta(0)/k_BT_c$\footnotemark[3]\\ \hline
Zn & 2.38 & 2.40 & 1.50 & 0.167 & 3.63 (3.20)\\
Cd & 0.51 & 1.50 & 1.20 & 0.245 & 3.70 (3.20)\\
Hg & 1.62 & 16.5 & 0.44 & 0.124 & 4.76 (4.60)\\
Al & 2.74 & 3.40 & 1.81 & 0.183 & 3.54 (3.30)\\
Ga & 7.92 & 3.30 & 1.51 & 0.096 & 3.51 (3.50)\\
In & 0.49 & 10.5 & 0.62 & 0.203 & 3.92 (3.60)\\
Tl & 0.46 & 7.35 & 0.48 & 0.164 & 3.92 (3.57)\\
Sn & 2.52 & 11.5 & 1.25 & 0.158 & 3.70 (3.50)\\
Pb & 1.39 & 27.3 & 0.60 & 0.160 & 4.56 (4.38)\\
V  & 27.9 & 16.0 & 1.20 & 0.068 & 3.95 (3.40)\\
Nb & 9.80 & 30.5 & 1.04 & 0.107 & 4.25 (3.80)\\
Ta & 6.72 & 14.0 & 0.90 & 0.095 & 3.95 (3.60)\\
\end{tabular}
\end{ruledtabular}
\footnotetext[1]{in $10^{-4}$eV} \footnotetext[2]{in $10^{-3}$}
\footnotetext[3]{experimental data bracketed}
\end{table}
The matrix element
\begin{equation}
\mathcal{M}_{\bf q} = \left[\frac{\hbar N}{2M\omega_{\bf
q}}\right]^{\frac{\scriptstyle 1}{\scriptstyle 2}}\!\!\int
_{\Omega}\psi _{\bf k'}^{*}({\bf r})\delta\mathcal{V}({\bf r})\psi
_{\bf k} ({\bf r})d{\bf r} \label{eq:matrix}
\end{equation}
depends on the atomic potential $\delta\mathcal V$, where $\psi$'s
are electron wave functions (spin absorbed into ${\bf k}$ and
${\bf k'}$), $M$ atomic mass, $N$ number of atoms in unit volume,
{\bf r} coordinates in real space and $\Omega$ the Wigner-Seitz
cell~\cite{Ziman, Mott}. Carbotte and Dynes utilized tabulated
data of the Heine-Abarenkov pseudopotential to estimate
$\delta\mathcal V$ and calculated the electric resistivity $\rho$
(another effect of the electron-phonon interaction) and energy gap
$\Delta$ (using the Eliashberg formalism)
separately~\cite{Carbotte1, Carbotte2}.  We use the Mott-Jones
formula [5] to link $\delta\mathcal V$ with $\rho$ and then use
this $\rho$ to express $\delta\mathcal V$: the Heine-Abarenkov
pseudopotential is no longer needed.  Letting ${\bf k'}$ in the
self-consistent equation [without terms in ${\bf k'}(\mathcal U)$]
run over a phonon sphere centered at the Fermi surface, we find:
\begin{eqnarray}
\Delta(\epsilon) &=& \Delta _0\left[\pi\int^{(4Z)^{-1/3}}_0
F^2(3.84Z^{1/3}\zeta)\zeta^3d\zeta\right]^{-1}\nonumber\\
&\times&\int^{\Xi}_{0}F^2(3.84Z^{1/3}\xi)\xi^2d\xi\nonumber\\
&\times&\frac{1}{2}\int^{1 + \xi}_{-1 + \xi}\frac{\Delta(\epsilon
+ 4\epsilon _F\xi x)}{E(\epsilon + 4\epsilon _F\xi x)}\;\frac{dx}
{\delta^2 - x^2} \label{eq:self_2}
\end{eqnarray}
Here $F$ is the overlap integral function~\cite{Ziman} and
\begin{equation}
\Delta _0 = \hbar e^2n\rho v^2/2k_BT_{\rho} \label{eq:Delta_0}
\end{equation}
measures the strength of electron-phonon interaction because it is
close to the value of $\sum_{\bf k'}V_{\bf kk'}$ (to a factor
$\sim$1), $\bf k'- k$ runs over the first phonon Brillouin zone,
$e$ and $n$ are the electron charge and density, $T_{\rho}$ the
temperature ($\sim$room temperature) when $\rho$ is measured, and
$v$ Debye sound velocity. Apparently strong electron-phonon
interaction arises from numerous free electrons (large $n$)
scattered frequently by atoms (large $\rho$) that move quickly to
facilitate pairing (large $v$).

We find $\Delta(\epsilon)$ through iteration.  The above
integration with respect to $x$ is of the Cauchy
type~\cite{Kuper}.  In first iteration we let $\Delta = \Delta_0$,
so that the Cauchy principal value has an analytical expression:
\begin{eqnarray}
&&\int\frac{\Delta _0}{\left[\Delta^2_0 + (\epsilon + 4\epsilon_F
\xi x)^2\right]^{1/2}}\;\frac{dx}{\delta^2 - x^2}\nonumber\\
&&= \frac{\Delta_0/2\delta}{\left[\Delta^2_0 + (\epsilon +
4\epsilon_F\xi\delta)^2\right]^{1/2}}\;\ln\frac{2A(x)}{|x -
\delta|}\nonumber\\
&& - \frac{\Delta _0/2\delta}{\left[\Delta^2_0 + (\epsilon -
4\epsilon _F\xi\delta)^2\right]^{1/2}}\;\ln\frac{2B(x)}{|x +
\delta|} + C \label{eq:analytic}
\end{eqnarray}
where $C$ is the integration constant,
\begin{eqnarray}
&& A(x)=\Delta^2_0 + (\epsilon + 4\epsilon _F\xi\delta)
(\epsilon + 4\epsilon _F\xi x)\nonumber\\
&& \;\;\;\;\;\;\;\; +\left[\Delta^2_0 + (\epsilon + 4\epsilon
_F\xi\delta)^2\right]^{1/2}\left[\Delta^2_0 + (\epsilon
+ 4\epsilon _F\xi x)^2\right]^{1/2}\nonumber\\
&& B(x) = \Delta^2_0 + (\epsilon - 4\epsilon _F\xi\delta)
(\epsilon + 4\epsilon _F\xi x)\nonumber\\
&& \;\;\;\;\;\;\;\; + \left[\Delta^2_0 + (\epsilon - 4\epsilon
_F\xi\delta)^2\right]^{1/2}\left[\Delta^2_0 + (\epsilon +
4\epsilon _F\xi x)^2\right]^{1/2}\nonumber
\end{eqnarray}
which is integrated numerically.  We vary $\Xi$, the upper limit
of the integration, until $\Delta(0)$ matches its observed value.
The result in FIG.~\ref{fig:fig2} (dashed curve) is virtually
identical to the full numerical solution (where the Cauchy
principal value is found numerically) in first iteration.  It can
be seen that prominent features of $\Delta(\epsilon)$ have already
emerged from first iteration: further iterations improve the
accuracy of solution but retain the physics.
\begin{figure}
\resizebox{8cm}{!}{\includegraphics{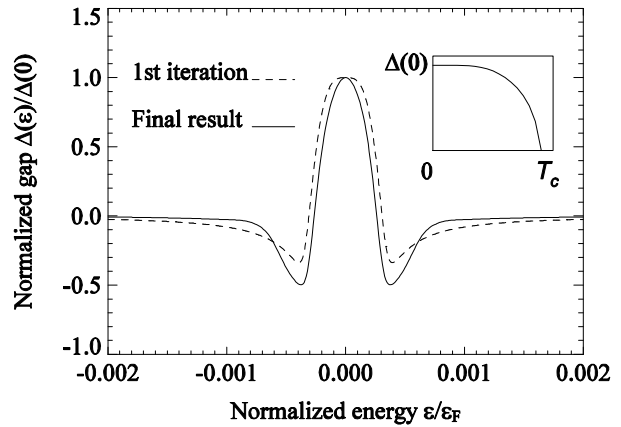}}
\caption{\label{fig:fig2} $\Delta(\epsilon)$ curve for Sn at $T =
0$.  The temperature curve of $\Delta$ at $\epsilon = 0$ is shown
in the insertion.}
\end{figure}

One such prominent feature is that $\Delta(\epsilon)$ is
structured: it has a peak flanked by two negative dips.  For Sn
the distance between the peak and one dip is $\sim 4\times
10^{-3}$ eV (FIG.~\ref{fig:fig2}), compared with $\hbar\omega_D =
1.7\times 10^{-3}$ eV: the pairing effect extends fairly deep into
the Fermi sea, in spite of the low $\omega_c$.  As a result, the
curve of $d\epsilon/dE$, which represents the tunnelling density
of states, also has a structure featuring a dip, similar to the
characteristic dip in the tunnelling experiment
data~\cite{Rowell}. Indeed $\omega(\epsilon)$ is also structured
in the numerical solutions for the Eliashberg
equation~\cite{Rowell, Scalapino}.

Another prominent feature is that $\omega_c/\omega_D << 1$ (0.148
on average for the 12 metals in Table~\ref{tab:table1}).  This is
somewhat puzzling because, assuming that $\Delta/E$ varies slowly
over the range of phonon frequencies, we can integrate the
self-consistent equation over the first phonon Brillouin zone and
find reasonable values of $\Delta$. Indeed, if we let $\Delta({\bf
k'})\approx\Delta({\bf k})$ and hence $E({\bf k'})\approx E({\bf
k})$, then we have $2E({\bf k})\approx\sum_{\bf k'}V_{\bf kk'}$,
or $\Delta(0)\approx\Delta_0/2$ (more or less true in
Table~\ref{tab:table1}). However $\Delta/E$ is by no means slowly
varying but has the shape of a sharp peak, which samples the
extremely large and positive values of $V_{\bf kk'}$, that would
have been cancelled by the equally large but negative values of
$V_{\bf kk'}$, had $\Delta/E$ been slowly varying. As a result
$\Delta(0)$ would exceed its experimental value vastly unless
$\omega_c/\omega_D << 1$.

In order to justify the values of $\omega_c/\omega_D$ in
Table~\ref{tab:table1} we solve the self-consistent equation at $T
> 0$, which is identical to its counterpart at $T = 0$, save an
additional factor $\tanh(E/2k_BT)$ in the integrand~\cite{BCS}.
Now $\Delta = \Delta(\epsilon, T)$ which is a function of $T$ also
found through iteration [$\Delta = \Delta(\epsilon)$ implies $T =
0$]. We let $T$ increase in small steps and use converged
$\Delta(\epsilon, T)$ to start iteration at the next $T$. We find
$T_c$ through a quadratic curve fit once $\Delta(0, T) <
0.01\Delta(0)$. Table~\ref{tab:table1} and FIG.~\ref{fig:fig3}
\begin{figure}
\resizebox{8cm}{!}{\includegraphics{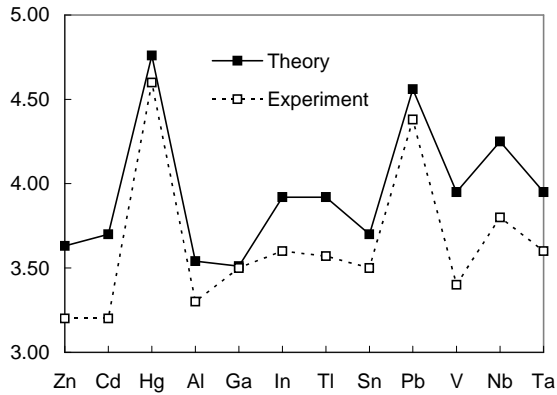}}
\caption{\label{fig:fig3} Theoretical and experimental values of
$2\Delta(0)/k_BT_c$.}
\end{figure}
show that $2\Delta(0)/k_BT_c$ matches experimental data reasonably
well (average error 8.9\%). In particular the large values 4.6 and
4.38 for Hg and Pb have been calculated fairly successfully.  If
we were to cut phonons off at $\omega_D$, then we would have high
$T_c$ or low (on average $\sim$2.5\% of measured values) with an
erroneous $2\Delta(0)/k_BT_c$ (e.g.\ 3.63 and 3.67 for Hg and Pb).

In conclusion the electron pairing cutoff frequency $\omega_c$ in
BCS theory may arise naturally.  Since $\omega_c/\omega_D << 1$
the 12 metals in Table~\ref{tab:table1}, Hg and Pb included, can
all be treated as weak-coupling superconductors, where the BCS
theory can be applied to yield e.g.\ correct values of
$2\Delta(0)/k_BT_c$: Landau's idea of a Fermi liquid covers
strong-coupling BCS superconductors. Strong electron-phonon
interaction may not be necessary for high $T_c$: Hg has a small
$\Delta_0$ but large energy gap, whereas V has a large $\Delta_0$
but roughly the same gap (see Table~\ref{tab:table1}). Our
conclusions are valid for BCS superconductors.  They may help us
to understand MgB$_2$ or other superconductors with umklapp
scattering~\cite{Monteverde, Schon, Struzhkin, Souma}.

\acknowledgements{The authors wish to thank Prof.\ R.\ Atkinson
for helpful comments.}

\end{document}